\documentclass[showpacs,aps,graphicx,twocolumn,]{revtex4}
\usepackage{graphicx}

\begin{document}

\title{Universal quantum gates for hybrid systems assisted by quantum dots inside double-sided optical
microcavities\footnote{Published in Phys. Rev. A \textbf{87}, 022305
(2013)}}

\author{Hai-Rui Wei  and Fu-Guo Deng\footnote{Corresponding author: fgdeng@bnu.edu.cn}}

\address{Department of Physics, Applied Optics Beijing Area Major
Laboratory, Beijing Normal University, Beijing 100875, China}

\date{\today }

\begin{abstract}

We present some deterministic schemes to construct universal quantum
gates, that is, controlled- NOT, three-qubit Toffoli, and Fredkin
gates, between flying photon qubits and stationary electron-spin
qubits assisted by quantum dots inside double-sided optical
microcavities. The control qubit of our gates is encoded on the
polarization of the moving single photon and the target qubits are
encoded on the confined electron spins in quantum dots   inside
optical microcavities. Our schemes for these universal quantum gates
on a hybrid system have some advantages. First, all the gates are
accomplished with a success probability of 100\% in principle.
Second, our schemes require no additional qubits.  Third, the
control qubits of the gates are easily manipulated and the target
qubits are perfect for storage and processing.  Fourth, the gates do
not require that the transmission for the uncoupled cavity is
balanceable with the reflectance for the coupled cavity, in order to
get a high fidelity. Fifth, the devices for the three universal
gates work in both the weak coupling and the strong coupling
regimes, and they are feasible in  experiment.

\end{abstract}


\maketitle

\section{Introduction}\label{Sec1}

Quantum logic gates lie at the heart of quantum information
processing. Two-qubit controlled-not (CNOT) gates  together with
single-qubit gates are sufficient for universal quantum computing
\cite{book,uni}. The optimal synthesis and the ``small-circuit"
structure for two-qubit systems have been well solved
\cite{2-qubit1,2-qubit2,2-qubit3,2-qubit4}, while the case for
multi-qubit systems is quite complex. Up to now, it has also been an
open question. That is, it is significant to seek a simpler scheme
for directly implementing multi-qubit gates. In the domain of
three-qubit gates, Toffoli and Fredkin gates have attracted much
attention  and both these two gates are  universal. Together with
Hadamard gates, they form a universal quantum computation
architecture \cite{Toffoli,Fredkin}. Moreover, they play an
important role in phase estimation \cite{book}, complex quantum
algorithms \cite{Shor}, error correction \cite{error}, and fault
tolerant quantum circuits \cite{fault}.

Quantum logic gates between flying photon qubits and stationary
(matter) qubits hold  great promise for quantum communication and
computing since photons are the perfect candidates for fast and
reliable long-distance communication because of their robustness
against decoherence, while the stationary qubits are suitable for
processor and local storage. In 2005, Liang and Li  \cite{swap}
chose atoms or ions as the matter qubit to discuss  the realization
of a two-qubit SWAP gate. One of the attractive candidates for a
stationary qubit is the electron spin in a GaAs-based or InAs-based
charged quantum dot (QD). The electron-spin coherence time of a
charged QD \cite{QD1,QD2,QD3,QD4,QD5,QD6} can be maintained for more
than 3 $\mu$s \cite{coher-time1,coher-time2} using spin echo
techniques, and the electron spin-relaxation time can be longer
($\sim ms$) \cite{relaxation1,relaxation2}. Moreover, it is
comparatively easy to incorporate a QD into a solid-state cavity,
and fast QD-spin cooling and manipulation have had some significant
progress \cite{cooling1,cooling2,manipulating1,manipulating2}. These
results indicate that the excess electron spin confined in a QD can
be used for a stable and scalable quantum computation. Based on a
singly charged QD inside an optical resonant cavity \cite{Hu1,Hu2},
in 2010 Bonato \emph{et al.}  \cite{CNOT} proposed a theoretical
scheme for a CNOT gate with the confined electron spin as the
control qubit and the polarization photon as the target qubit. This
QD-cavity system has been used for a two-photon Bell-state analyzer,
entanglement generators, teleportation, entanglement swapping,
quantum repeaters, entanglement purification and concentration, and
hyperentangled-Bell-state analysis
\cite{Hu1,Hu2,CNOT,Hu3,Hu4,Appli1,Appli2,Appli3,Renbaocang}.

Different from the work by Bonato \emph{et al.} \cite{CNOT} in which
they presented a scheme for a CNOT gate with a confined electron
spin in a QD as the control qubit and a flying photon  as the target
qubit, we  first present a deterministic  scheme for constructing a
CNOT gate on a hybrid system with the flying photon as the control
qubit and the excess electron in a QD as the target qubit. Also, we
propose two deterministic schemes for constructing the Toffoli and
Fredkin gates on a three-qubit hybrid system. In our work, the
control qubits of our universal gates are encoded on the moving
photon qubit (i.e., the two polarization states of a single photon,
denoted by $|R\rangle$ and $|L\rangle$), while the target qubit is
encoded on the spin of the excess electron confined in a QD inside
an optical microcavity (denoted by $|\uparrow\rangle$ and
$|\downarrow\rangle$). These three schemes for the universal gates
require no additional qubits, and they only need some linear optical
elements besides the nonlinear interaction between the moving photon
and the electron in a QD inside an optical microcavity. It is worth
pointing out that these gates are robust because they do not require
that the transmission for the uncoupled cavity is balanceable with
the reflectance for the coupled cavity in order to get a high
fidelity, different from the hybrid gates which are encoded on the
spin confined in a QD inside a single-sided cavity \cite{Hu2}. The
fidelities and the efficiencies of our gates are discussed. A high
fidelity and a high efficiency can be achieved in both the strong
and the weak regimes, and our devices are feasible with current
experimental technology.

This paper is organized as follows: In Sec.\ref{Sec2}, we briefly
review  a singly charged QD in a double-sided optical microcavity.
In Sec.\ref{Sec3}, we propose a deterministic scheme for
constructing a two-qubit CNOT gate between a flying photon (the
control qubit) and a stationary electron (the target qubit) confined
in a QD. The fundamental three-qubit gates, Toffoli (control-CNOT)
and Fredkin (control-SWAP) gates, on a three-qubit hybrid system are
constructed in Secs.\ref{Sec4} and  \ref{Sec5}, respectively. The
fidelity, the efficiency, and the experimental feasibility of our
schemes for hybrid quantum gates are discussed in Sec.\ref{Sec6}.
Some discussions and a summary are given in Sec.\ref{Sec7}.

\section{A singly charged QD in a double-sided microcavity}
\label{Sec2}

\begin{figure}[!h]
\begin{center}
\includegraphics[width=6.4 cm,angle=0]{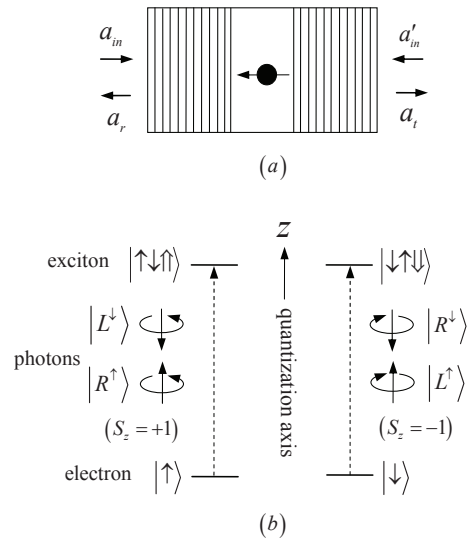}
\caption{(a) A schematic diagram for a singly charged QD inside a
double-sided optical microcavity. (b) Schematic description of the
relevant exciton energy levels and the spin selection rules for
optical transition of negatively charged exciton. The symbols
$\Uparrow$ ($\Downarrow$) and $\uparrow$ ($\downarrow$) represent a
hole and an excess electron with $z$-direction spin projections
$|+\frac{3}{2}\rangle$ ($|-\frac{3}{2}\rangle$) and
$|+\frac{1}{2}\rangle$ ($|-\frac{1}{2}\rangle$), respectively.
$|R^\uparrow\rangle$ ($|R^\downarrow\rangle$) denotes a
right-circularly polarized photon propagating along (against) the
normal direction of the cavity $z$ axis (the quantization axis) and
$|L^\uparrow\rangle$ ($|L^\downarrow\rangle$) denotes a
left-circularly polarized photon propagating along (against) the $z$
axis.} \label{Fig1}
\end{center}
\end{figure}

In 2009, Hu \emph{et al.} \cite{Hu2} proposed a  double-side
QD-cavity system, which can be used for quantum computation, quantum
communication, and quantum storage. In this appealing system, a
singly charged electron  In(Ga)As QD or a GaAs interface QD is
embedded in an optical resonant double-sided microcavity with two
mirrors partially reflective in the top and the bottom. The excess
electron-spin qubit in the QD promises  scalable  quantum
computation in solid-state systems,  and it interacts with the
cavity mode through the addition of a  negatively  charged  exciton
($X^-$) that consists of two electrons and one hole \cite{exciton1}
created by excitation. The exciton determines the rules of the
spin-dependent optical transitions (depicted in Fig.\ref{Fig1})
\cite{exciton2}. In this work, we consider the dipole resonance with
the cavity mode and the input photon. The rules of the input states
changing under the interaction of the photon and the cavity can be
described as follows:
\begin{eqnarray}   \label{eq.1}
|R^\uparrow\uparrow\rangle &\rightarrow&
|L^\downarrow\uparrow\rangle,\;\;\;\;\;\;\;\;\;
|L^\uparrow\uparrow\rangle \;\rightarrow\; -|L^\uparrow\uparrow\rangle, \nonumber\\
|R^\downarrow\uparrow\rangle &\rightarrow&
-|R^\downarrow\uparrow\rangle,\;\;\;\;\;\;
|L^\downarrow\uparrow\rangle \;\rightarrow\; |R^\uparrow\uparrow\rangle,\nonumber\\
|R^\uparrow\downarrow\rangle &\rightarrow&
-|R^\uparrow\downarrow\rangle,\;\;\;\;\;\;
|L^\uparrow\downarrow\rangle \;\rightarrow\; |R^\downarrow\downarrow\rangle,\nonumber\\
|R^\downarrow\downarrow\rangle &\rightarrow&
|L^\uparrow\downarrow\rangle,\;\;\;\;\;\;\;\;\;
|L^\downarrow\downarrow\rangle \;\rightarrow\;
-|L^\downarrow\downarrow\rangle.
\end{eqnarray}

There are two kinds of optical transitions between the electron and
the exciton $X^-$, one involving the  photon with the spin $s_z=+1$
and the other involving the photon with the spin $s_z=-1$. For a
photon with $s_z=+1$ ($|R^\uparrow\rangle$ or
$|L^\downarrow\rangle$), if the excess electron is in the spin state
$|\uparrow\rangle$, it couples to the dipole and will be reflected
by the cavity, and both the polarization and the propagation
direction of the photon will be flipped. If the excess electron spin
is in the state $|\downarrow\rangle$,  the photon in the
polarization state $|R^\uparrow\rangle$ or $|L^\downarrow\rangle$
will not couple to the dipole, and it will transmit the cavity and
acquire a $\pi$ mod $2\pi$ phase shift relative to a reflected
photon.  In the same way, for the photon in the state
$|R^\downarrow\rangle$ or $|L^\uparrow\rangle$ ($s_z=-1$), if the
excess electron spin is in the state $|\uparrow\rangle$, it will
transmit the cavity. If the excess electron spin is in the state
$|\downarrow\rangle$, it will be reflected by the cavity. That is,
this structure ( $X^--$cavity system) acts as an entanglement beam
splitter. It splits an initial product state of the system composed
of an injecting photon and an electron spin into two entangled
states via the transmission and the reflection of the photon in a
deterministic way. Compared with a single-sided QD-cavity system,
the doubled-sided unit   easily reaches a large phase difference
($\pi$) between the uncoupled cavity and the coupled cavity
\cite{Hu2}. The device based on double-sided units is robust
\cite{Hu2}. In the following, we investigate the construction of the
universal hybrid quantum gates, that is, CNOT, Toffoli, and Fredkin
gates, which take electron-spin qubits confined in QDs as the target
qubits and a flying photon as the control qubit.

\section{ CNOT gate on a two-qubit hybrid system with a flying photon as the control qubit}
\label{Sec3}

\begin{figure}[!h]
\begin{center}
\includegraphics[width=5.2cm,angle=0]{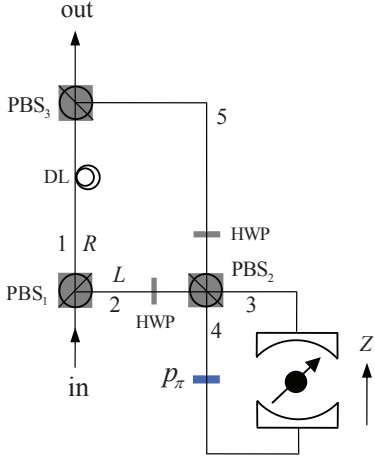}
\caption{(Color online) The quantum circuit for constructing a
deterministic CNOT gate with a flying photon polarization as the
control qubit and a confined electron spin as the target qubit.
PBS$_i$ ($i=1,2,3$) is a polarizing beam splitter in the circular
basis, which transmits the right-circular polarization photon
$|R\rangle$  and reflects the left-circular polarization photon
$|L\rangle$, respectively. HWP represents a half-wave plate which is
set to $22.5^\circ$ to induce the Hadamard transformations on the
polarizations of photons. $P_\pi$ is a phase shifter that
contributes a $\pi$ phase shift to the photon passing through it. DL
is the time-delay device for making the photons from   modes 1 and 5
reach PBS$\,_3$ simultaneously. Before and after the photon
interacts with the electron spin in the QD inside the double-side
optical microcavity, a Hadamard ($H_{e}$) operation is performed on
the electron spin by using a $\pi/2$ microwave pulse or an optical
pulse.} \label{Fig2}
\end{center}
\end{figure}

The principle for a CNOT gate, which flips the target electron-spin
qubit if the control photon polarization qubit is in the state
$|L\rangle$, is depicted in Fig.\ref{Fig2}. The flying photon  $p$
and the excess electron $e$ in a QD are prepared in arbitrary
superposition sates
$|\psi\rangle_p=\alpha_c|R\rangle+\beta_c|L\rangle$ and
$|\psi\rangle_e=\alpha_t|\uparrow\rangle+\beta_t|\downarrow\rangle$
(here $|\alpha_c|^2 + |\beta_c|^2 =  |\alpha_t|^2 + |\beta_t|^2=1$),
respectively. The subscripts $p$ and $e$ represent the photon and
the electron, respectively. The CNOT gate can be constructed with
the steps shown in Fig.\ref{Fig2}.

First, the injecting photon passes through PBS$_1$ which transmits
the photon in the polarization state $|R\rangle$ and reflects the
photon in the state  $|L\rangle$. That is, PBS$_1$ splits the photon
into two wave-packets. The part in the state $|L\rangle$ is injected
into the cavity and interacts with the QD, while the part in the
state $|R\rangle$ transmits  PBS$_1$ and does not interact with the
cavity. After the control photon passes through PBS$_1$, the state
of the whole system composed of a photon and an electron is changed
from $|\Psi_0\rangle$ to $\vert \Psi_1\rangle$. Here
\begin{eqnarray}                                  \label{eq.2}
\vert \Psi_0\rangle &=& |\psi\rangle_p \otimes |\psi\rangle_e\nonumber\\
&=& (\alpha_c|R\rangle+\beta_c|L\rangle) \otimes(\alpha_t|\uparrow\rangle+\beta_t|\downarrow\rangle),  \\
\vert \Psi_1\rangle &=&
      \alpha_c\alpha_t|R\rangle|\uparrow\rangle
      +\alpha_c\beta_t|R\rangle|\downarrow\rangle \nonumber\\
  &+&
  \beta_c\alpha_t|L^\uparrow\rangle|\uparrow\rangle
     +\beta_c\beta_t|L^\uparrow\rangle|\downarrow\rangle.
\end{eqnarray}

Before the photon coming from the spatial mode $2$ passes through
PBS$_2$, a Hadamard ($H_{p}$) operation is performed on it with a
half-wave plate (HWP) which is used to complete the transformations
\begin{eqnarray}                                      \label{eq.4}
|R\rangle \;\rightarrow\; \frac{1}{\sqrt{2}}(|R\rangle+|L\rangle),
\;\;\; |L\rangle \;\rightarrow\;
\frac{1}{\sqrt{2}}(|R\rangle-|L\rangle),
\end{eqnarray}
and   a Hadamard ($H_{e}$) operation (e.g., using a $\pi/2$
microwave pulse or optical pulse
\cite{manipulating1,manipulating2})is also performed on the electron
to complete the transformations
\begin{eqnarray}   \label{eq.5}
|\uparrow\rangle &\rightarrow & |\rightarrow \rangle \equiv
\frac{1}{\sqrt{2}}(|\uparrow\rangle+|\downarrow\rangle),\nonumber\\
|\downarrow\rangle & \rightarrow &  |\leftarrow \rangle \equiv
\frac{1}{\sqrt{2}}(|\uparrow\rangle-|\downarrow\rangle).
\end{eqnarray}
PBS$_2$ will lead the photon to  paths $3$ and $4$ when the photon
is in the states $\vert R\rangle$ and $\vert L\rangle$,
respectively. When the photon passes through  path $4$, it takes a
phase shift $\pi$ (i.e., $\vert L\rangle\rightarrow -\vert L\rangle$
and $\vert R\rangle\rightarrow -\vert R\rangle$ when the photon
passes through the device $P_\pi$). That is, before the photon
interacts with the QD inside the optical microcavity, the state of
the photon-electron system  becomes
\begin{eqnarray}                                  \label{eq.6}
\vert\Psi_2\rangle &=& \alpha_c|R\rangle(\alpha_t|\rightarrow\rangle  + \beta_t |\leftarrow\rangle) \nonumber\\
  & +&  \frac{1}{2}\beta_c\alpha_t(|R^\downarrow\rangle + |L^\uparrow\rangle)(|\uparrow\rangle + |\downarrow\rangle)\nonumber\\
  & +&  \frac{1}{2} \beta_c\beta_t (|R^\downarrow\rangle + |L^\uparrow\rangle)(|\uparrow\rangle - |\downarrow\rangle).
\end{eqnarray}
The nonlinear  interaction between the photon and the electron in
the QD assisted by the optical microcavity makes the state of the
system be changed as
\begin{eqnarray}                                  \label{eq.7}
\vert\Psi_3\rangle &=& \alpha_c|R\rangle(\alpha_t|\rightarrow\rangle  + \beta_t |\leftarrow\rangle) \nonumber\\
  & +&  \frac{1}{2}\beta_c\alpha_t(-|R^\downarrow\rangle- |L^\uparrow\rangle)(|\uparrow\rangle - |\downarrow\rangle)\nonumber\\
  & +&  \frac{1}{2} \beta_c\beta_t (-|R^\downarrow\rangle - |L^\uparrow\rangle)(|\uparrow\rangle + |\downarrow\rangle).
\end{eqnarray}
When the photon $p$ is in the state $|R^\downarrow\rangle$, it
passes through the phase shifter $P_\pi$ and reaches  PBS$_2$ from
  path $4$. When the photon $p$ is in the state
$|L^\uparrow\rangle$, it does not take a phase shift and reaches
PBS$_2$  from  path $3$. Whether the photon passes through the path
$4$ or  path $3$, it is emitted from  path $5$ after an $H_p$
operation is performed on it. Also, an $H_e$ operation  is performed
on the electron spin. After the photon passes through PBS$_3$, the
state of the system becomes
\begin{eqnarray}                                  \label{eq.8}
\vert\Psi_4\rangle &=& \alpha_c|R\rangle (\alpha_t|\uparrow\rangle
+\beta_t|\downarrow\rangle) \nonumber\\
&+& \beta_c|L\rangle(\alpha_t|\downarrow\rangle +
\beta_t|\uparrow\rangle).
\end{eqnarray}

From Eq.(\ref{eq.8}), one can see that the state of the electron
(the target qubit) is flipped when the photon (the control qubit) is
in the state $\vert L\rangle$, while it does not change when the
photon is in the state $\vert R\rangle$, compared to the original
state of the two-qubit hybrid system shown in Eq.(\ref{eq.2}). That
is, the quantum circuit shown in Fig.\ref{Fig2} can be used to
construct a deterministic CNOT gate with a success probability of
100\% in principle, by using the photon as the control qubit and the
electron spin as the target qubit.

\begin{figure}[!h]
\begin{center}
\includegraphics[width=5.5cm,angle=0]{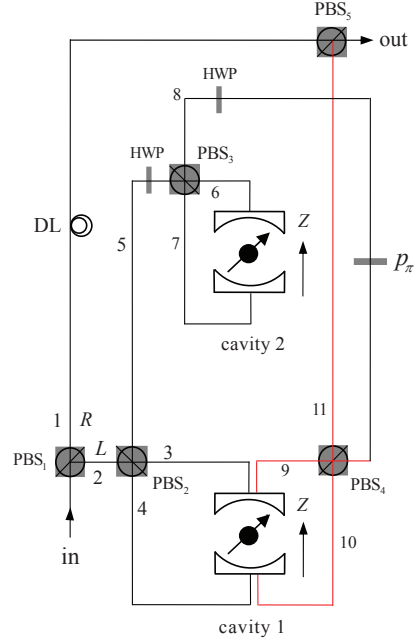}
\caption{ (Color online) Scheme for implementing a three-qubit
Toffoli gate with a flying photon polarization and a confined
electron-spin qubit as the two control qubits and another confined
electron-spin qubit as the target qubit. } \label{Fig3}
\end{center}
\end{figure}

\section{Toffoli gate on a three-qubit Hybrid system}\label{Sec4}

A three-qubit Toffoli gate is used to perform a NOT operation on a
target qubit or not, depending on the states of the two control
qubits \cite{Fredkin}. In the optimal scheme for decomposing a
three-qubit gate, it requires at least six CNOT gates for
implementing a Toffoli gate \cite{cost}, which increases the
difficulty of its implementation by using the fundamental two-qubit
gates and one-qubit gates. Therefore, it is desirable to seek a
simpler scheme for directly implementing the three-qubit Toffoli
gate.

Our device for implementing a deterministic Toffoli gate on a photon
qubit and two electron-spin qubits is shown in Fig.\ref{Fig3}. When
the control photon qubit (the flying photon) and the control
electron-spin qubit (the electron in cavity 1) are in the states
$|L\rangle$ and $|\downarrow\rangle$, respectively, the state of the
target electron-spin qubit  (the electron in cavity 2) is flipped;
otherwise, the state of the target electron-spin qubit does not
change. We describe this principle in detail as follows.

Suppose that the flying photon qubit  is prepared in an arbitrary
superposition state
\begin{eqnarray}  \label{eq.9}
|\psi\rangle_p=\alpha_p|R\rangle+\beta_p|L\rangle,
\end{eqnarray}
and each of two independent excess electrons in cavities 1 and 2 is
prepared in an arbitrary state as
\begin{eqnarray}  \label{eq.10}
|\psi\rangle_{e_1}=\alpha_{e_1}|\uparrow\rangle_{1}+\beta_{e_1}|\downarrow\rangle_{1}, \nonumber\\
|\psi\rangle_{e_2}=\alpha_{e_2}|\uparrow\rangle_{2}+\beta_{e_2}|\downarrow\rangle_{2}.
\end{eqnarray}
Here the subscript $1$ ($2$) represents the electron in cavity 1 (2)
and
\begin{eqnarray}  \label{eq.11}
|\alpha_p|^2 + |\beta_p|^2 =  |\alpha_{e_1}|^2 + |\beta_{e_1}|^2=
|\alpha_{e_2}|^2 + |\beta_{e_2}|^2 = 1.
\end{eqnarray}
We first inject the flying photon $p$ from PBS$_1$ which splits the
photon into two wave-packets, the part in the state $|R\rangle$ and
that in $|L\rangle$. When the photon is in the state  $|L\rangle$,
it passes through PBS$_2$ and is injected into  cavity $1$.  When
the photon is in the state $|R\rangle$, it transmits  PBS$_1$ and
does not interact with the cavity.  After the photon passes through
PBS$_1$, the state of the whole system composed of a flying photon
and two stationary electrons is changed from $\vert \Phi_0\rangle$
to $\vert \Phi_1\rangle$. Here
\begin{eqnarray}                                  \label{eq.12}
\vert \Phi_0\rangle &=& |\psi\rangle_p
\otimes|\psi\rangle_{e_1}\otimes |\psi \rangle_{e_2} \nonumber\\
&=&
\alpha_p\alpha_{e_1}|R\rangle|\uparrow\rangle_{1}(\alpha_{e_2}|\uparrow\rangle
+ \beta_{e_2} |\downarrow\rangle)_{2} \nonumber\\
&+&
\alpha_p\beta_{e_1}|R\rangle|\downarrow\rangle_{1}(\alpha_{e_2}|\uparrow\rangle
+ \beta_{e_2} |\downarrow\rangle)_{2}\nonumber\\
&+&\beta_p\alpha_{e_1}
|L\rangle|\uparrow\rangle_{1}(\alpha_{e_2}|\uparrow\rangle
+ \beta_{e_2} |\downarrow\rangle)_{2}\nonumber\\
&+&\beta_p\beta_{e_1}|L\rangle|\downarrow\rangle_{1}(\alpha_{e_2}|\uparrow\rangle
+ \beta_{e_2} |\downarrow\rangle)_{2},\\
\nonumber\\
\vert \Phi_1\rangle &=& \alpha_p\alpha_{e_1}\alpha_{e_2}|R
\rangle|\uparrow  \uparrow\rangle_{12}+
\alpha_p\alpha_{e_1}\beta_{e_2}|R \rangle|\uparrow
\downarrow\rangle_{12} \nonumber\\
&+& \alpha_p\beta_{e_1}\alpha_{e_2}|R \rangle|\downarrow
\uparrow\rangle_{12} + \alpha_p\beta_{e_1}\beta_{e_2}|R
\rangle|\downarrow \downarrow\rangle_{12} \nonumber\\
&+& \beta_p\alpha_{e_1}\alpha_{e_2}|L^\uparrow\rangle|\uparrow
\uparrow\rangle_{12}
+\beta_p\alpha_{e_1}\beta_{e_2}|L^\uparrow\rangle|\uparrow
\downarrow\rangle_{12} \nonumber\\
&+& \beta_p\beta_{e_1}\alpha_{e_2}|L^\uparrow\rangle|\downarrow
\uparrow\rangle_{12}
+\beta_p\beta_{e_1}\beta_{e_2}|L^\uparrow\rangle|\downarrow \downarrow\rangle_{12}.\nonumber\\
\end{eqnarray}

The photon in the state $|L^\uparrow\rangle$ is injected into cavity
1 and interacts with the QD inside the microcavity. This nonlinear
interaction makes the state of the system be changed as
\begin{eqnarray}            \label{eq.14}
|\Phi_2\rangle &=& \alpha_p\alpha_{e_1}\alpha_{e_2}|R
\rangle|\uparrow\uparrow\rangle_{12}+
\alpha_p\alpha_{e_1}\beta_{e_2}|R\rangle|\uparrow\downarrow\rangle_{12}\nonumber\\
&+&\alpha_p\beta_{e_1}\alpha_{e_2}|R\rangle|\downarrow\uparrow\rangle_{12}
+\alpha_p\beta_{e_1}\beta_{e_2}|R\rangle|\downarrow\downarrow\rangle_{12}\nonumber\\
&-&\beta_p\alpha_{e_1}\alpha_{e_2}|L^\uparrow\rangle|\uparrow\uparrow\rangle_{12}
-\beta_p\alpha_{e_1}\beta_{e_2}|L^\uparrow\rangle|\uparrow\downarrow\rangle_{12}\nonumber\\
&+&\beta_p\beta_{e_1}\alpha_{e_2}|R^\downarrow\rangle|\downarrow\uparrow\rangle_{12}
+\beta_p\beta_{e_1}\beta_{e_2}|R^\downarrow\rangle|\downarrow\downarrow\rangle_{12}.\nonumber\\
\end{eqnarray}
Whether the photon is in the state $|L^\uparrow\rangle$ from  path 3
or the photon is in the state $|R^\downarrow\rangle$ from  path 4,
it will be emerged in  spatial mode 5 by PBS$_2$. Before the photon
from  path 5 passes through PBS$_3$ and then is injected into cavity
2, an $H_p$ operation is performed on it (i.e., passing through the
HWP), and an $H_e$ operation is also performed on the electron in
cavity 2. That is, before the photon interacts with the QD in cavity
2, the state of the system becomes
\begin{eqnarray}            \label{eq.15}
|\Phi_3\rangle &=&
\alpha_p\alpha_{e_1}\alpha_{e_2}|R\rangle|\uparrow\rightarrow\rangle_{12}+
\alpha_p\alpha_{e_1}\beta_{e_2}|R\rangle|\uparrow\leftarrow\rangle_{12}\nonumber\\
&+&\alpha_p\beta_{e_1}\alpha_{e_2}|R\rangle|\downarrow\rightarrow\rangle_{12}
+\alpha_p\beta_{e_1}\beta_{e_2}|R\rangle|\downarrow\leftarrow\rangle_{12}\nonumber\\
&-&\frac{1}{2}\beta_p\alpha_{e_1}\alpha_{e_2}(|R^\downarrow\rangle-|L^\uparrow\rangle)|\uparrow\rangle_{1}(|\uparrow
\rangle
+|\downarrow\rangle)_{2}\nonumber\\
&-&\frac{1}{2}\beta_p\alpha_{e_1}\beta_{e_2}(|R^\downarrow\rangle-|L^\uparrow\rangle)|\uparrow\rangle_{1}(|\uparrow
\rangle
-|\downarrow\rangle)_{2}\nonumber\\
&+&\frac{1}{2}\beta_p\beta_{e_1}\alpha_{e_2}(|R^\downarrow\rangle+|L^\uparrow\rangle)|\downarrow\rangle_{1}(|\uparrow
\rangle
+|\downarrow\rangle)_{2}\nonumber\\
&+&\frac{1}{2}\beta_p\beta_{e_1}\beta_{e_2}(|R^\downarrow\rangle+|L^\uparrow\rangle|\downarrow\rangle_{1}(|\uparrow
\rangle
-|\downarrow\rangle)_{2}.
\end{eqnarray}

The nonlinear interaction between the photon and the QD inside
cavity  2  makes  the state given by Eq.(\ref{eq.15}) become
\begin{eqnarray}            \label{eq.16}
|\Phi_4\rangle &=&
\alpha_p\alpha_{e_1}\alpha_{e_2}|R\rangle|\uparrow\rightarrow\rangle_{12}+
\alpha_p\alpha_{e_1}\beta_{e_2}|R\rangle|\uparrow\leftarrow\rangle_{12}\nonumber\\
&+&\alpha_p\beta_{e_1}\alpha_{e_2}|R\rangle|\downarrow\rightarrow\rangle_{12}
+\alpha_p\beta_{e_1}\beta_{e_2}|R\rangle|\downarrow\leftarrow\rangle_{12}\nonumber\\
&+&\frac{1}{2}\beta_p\alpha_{e_1}\alpha_{e_2}(|R^\downarrow\rangle-|L^\uparrow\rangle)|\uparrow\rangle_{1}(|\uparrow\rangle
+|\downarrow\rangle)_{2}\nonumber\\
&+&\frac{1}{2}\beta_p\alpha_{e_1}\beta_{e_2}(|R^\downarrow\rangle-|L^\uparrow\rangle)|\uparrow\rangle_{1}(|\uparrow\rangle
-|\downarrow\rangle)_{2}\nonumber\\
&-&\frac{1}{2}\beta_p\beta_{e_1}\alpha_{e_2}(|R^\downarrow\rangle+|L^\uparrow\rangle)|\downarrow\rangle_{1}(|\uparrow\rangle
-|\downarrow\rangle)_{2}\nonumber\\
&-&\frac{1}{2}\beta_p\beta_{e_1}\beta_{e_2}(|R^\downarrow\rangle+|L^\uparrow\rangle)|\downarrow\rangle_{1}(|\uparrow\rangle
+|\downarrow\rangle)_{2}.
\end{eqnarray}
Next, an $H_p$ operation (i.e., the HWP in  path 8) is performed on
the photon in the state $|R^\downarrow\rangle$ or the photon in the
state $|L^\uparrow\rangle$ coming from paths 7 and   6,
respectively, and  an  $H_e$  operation is also performed on the
electron in cavity 2. After these two operations, the state of the
system becomes
\begin{eqnarray}             \label{eq.17}
|\Phi_5\rangle &=&
\alpha_p\alpha_{e_1}\alpha_{e_2}|R\rangle|\uparrow\uparrow\rangle_{12}+
\alpha_p\alpha_{e_1}\beta_{e_2}|R\rangle|\uparrow\downarrow\rangle_{12}\nonumber\\
&+&\alpha_p\beta_{e_1}\alpha_{e_2}|R\rangle|\downarrow\uparrow\rangle_{12}
+\alpha_p\beta_{e_1}\beta_{e_2}|R\rangle|\downarrow\downarrow\rangle_{12}\nonumber\\
&+&\beta_p\alpha_{e_1}\alpha_{e_2}|L^\uparrow\rangle|\uparrow\uparrow\rangle_{12}
+\beta_p\alpha_{e_1}\beta_{e_2}|L^\uparrow\rangle|\uparrow\downarrow\rangle_{12}\nonumber\\
&-&\beta_p\beta_{e_1}\alpha_{e_2}|R^\downarrow\rangle|\downarrow\downarrow\rangle_{12}
-\beta_p\beta_{e_1}\beta_{e_2}|R^\downarrow\rangle|\downarrow\uparrow\rangle_{12}.\nonumber\\
\end{eqnarray}

After the photon passes through the HWP in  spatial mode  8, it is
led back to cavity 1.  Before the photon reaches PBS$_4$, it passes
through a phase shifter $P_\pi$. $P_\pi$ makes
$|R^\downarrow\rangle$ and $|L^\uparrow\rangle$ in Eq.(\ref{eq.17})
become $-|R^\downarrow\rangle$ and $-|L^\uparrow\rangle$,
respectively. When the photon passes through cavity 1, the nonlinear
interaction between the photon and the QD in cavity 1 induces the
state of the system to be
\begin{eqnarray}   \label{eq.18}
|\Phi_6\rangle &=&
\alpha_p\alpha_{e_1}\alpha_{e_2}|R\rangle|\uparrow\uparrow\rangle_{12}+
\alpha_p\alpha_{e_1}\beta_{e_2}|R\rangle|\uparrow\downarrow\rangle_{12}\nonumber\\
&+&\alpha_p\beta_{e_1}\alpha_{e_2}|R\rangle|\downarrow\uparrow\rangle_{12}
+\alpha_p\beta_{e_1}\beta_{e_2}|R\rangle|\downarrow\downarrow\rangle_{12}\nonumber\\
&+&\beta_p\alpha_{e_1}\alpha_{e_2}|L^\uparrow\rangle|\uparrow\uparrow\rangle_{12}
+\beta_p\alpha_{e_1}\beta_{e_2}|L^\uparrow\rangle|\uparrow\downarrow\rangle_{12}\nonumber\\
&+&\beta_p\beta_{e_1}\alpha_{e_2}|L^\uparrow\rangle|\downarrow\downarrow\rangle_{12}
+\beta_p\beta_{e_1}\beta_{e_2}|L^\uparrow\rangle|\downarrow\uparrow\rangle_{12}.\nonumber\\
\end{eqnarray}
When the photon in the state $|L^\uparrow\rangle$ coming from cavity
1 passes through  PBS$_4$ again, it is reflected to PBS$_5$. When
the two wavepackets from spatial modes $1$ and $11$ pass through
PBS$_5$  simultaneously, the system composed of the photon and the
two electrons is in the state
\begin{eqnarray}   \label{eq.19}
|\Phi_7\rangle &=&
\alpha_p\alpha_{e_1}|R\rangle|\uparrow\rangle_{1}(\alpha_{e_2}|\uparrow\rangle
+ \beta_{e_2} |\downarrow\rangle)_{2} \nonumber\\
&+&
\alpha_p\beta_{e_1}|R\rangle|\downarrow\rangle_{1}(\alpha_{e_2}|\uparrow\rangle
+ \beta_{e_2} |\downarrow\rangle)_{2}\nonumber\\
&+&\beta_p\alpha_{e_1}
|L\rangle|\uparrow\rangle_{1}(\alpha_{e_2}|\uparrow\rangle
+ \beta_{e_2} |\downarrow\rangle)_{2}\nonumber\\
&+&\beta_p\beta_{e_1}|L\rangle|\downarrow\rangle_{1}(\alpha_{e_2}|\downarrow\rangle
+ \beta_{e_2} |\uparrow\rangle)_{2}.
\end{eqnarray}

From Eq.(\ref{eq.19}), one can see that the state of the target
qubit (i.e., the electron spin $e_2$ in cavity 2) is flipped when
the flying photon $p$ is in the state $\vert L\rangle$ and the
control electron-spin qubit $e_1$ is in the state $\vert
\downarrow\rangle_{1}$, compared to the state shown in
Eq.(\ref{eq.12}). Otherwise, it does not change. That is, the
quantum circuit shown in Fig.\ref{Fig3} can be used to construct a
Toffoli gate on a three-qubit hybrid system, by using the flying
photon and the electron confined in the first optical microcavity as
the two control qubits and using the electron confined in the second
microcavity as the target qubit. In principle, this Toffoli gate has
a success probability of 100\%. It is a deterministic three-qubit
gate.

\begin{figure}[!h]
\begin{center}
\includegraphics[width=5.5 cm,angle=0]{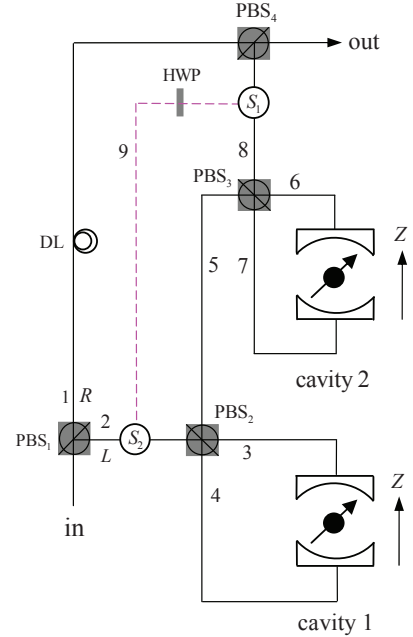}
\caption{(Color online) Schematic setup for a deterministic
three-qubit Fredkin gate with a flying photon polarization as the
control qubit and two confined electron spins as the target qubits.
$S_1$ and $S_2$ are two  optical switches. } \label{Fig4}
\end{center}
\end{figure}

\section{Fredkin gate on a three-qubit Hybrid system}\label{Sec5}

The matrix representation of the three-qubit Fredkin gate can be
written as
\begin{eqnarray}                                      \label{20}
U_F=diag \left\{I_4, \left(\begin{array}{cccc}
1&0&0&0\\
0&0&1&0\\
0&1&0&0\\
0&0&0&1\\
\end{array}\right)\right\},
\end{eqnarray}
in the   basis $\{$$|R\rangle|\uparrow\uparrow\rangle_{12}$,
$|R\rangle|\uparrow\downarrow\rangle_{12}$,
$|R\rangle|\downarrow\uparrow\rangle_{12}$,
$|R\rangle|\downarrow\downarrow\rangle_{12}$,
$|L\rangle|\uparrow\uparrow\rangle_{12}$,
$|L\rangle|\uparrow\downarrow \rangle_{12}$,
$|L\rangle|\downarrow\uparrow \rangle_{12}$,
$|L\rangle|\downarrow\downarrow \rangle_{12}$$\}$. Here $I_4$ is the
 four-dimensional unit matrix. That is, the gate implements a swap
operation on two stationary electron-spin qubits when the control
qubit (the flying photon $p$) is in the state $|L\rangle$.
Otherwise, it does nothing.

Now, let us  discuss how to construct this thee-qubit gate. Its
schematic setup is shown in Fig.\ref{Fig4}. Suppose that the initial
state of the flying photon and the two stationary electrons confined
in the two QDs inside two mocrocavities (cavity 1 and cavity 2) are
$|\psi\rangle_p=\alpha_c|R\rangle+\beta_c|L\rangle$ and
$|\psi\rangle_{e_i}=\alpha_{t_i}\vert\uparrow\rangle_{i}+\beta_{t_i}\vert\downarrow\rangle_{i}$
(here $i=1,2$, and $|\alpha_c|^2 + |\beta_c|^2 =  |\alpha_{t_i}|^2 +
|\beta_{t_i}|^2=1$), respectively.  That is, the initial state of
the whole system composed a photon and two electron spins can be
written as
\begin{eqnarray}                                  \label{eq.21}
\vert \Omega_0\rangle &=& |\psi\rangle_p \otimes|\psi\rangle_{e_1}\otimes |\psi\rangle_{e_2},\nonumber\\
&=&\vert \Omega_0'\rangle+\vert \Omega_0''\rangle.
\end{eqnarray}
Here
\begin{eqnarray}                          \label{eq.22}
\vert
\Omega_0'\rangle&=&\alpha_c\alpha_{t_1}\alpha_{t_2}|R\rangle|\uparrow\uparrow\rangle_{12}+
\alpha_c\alpha_{t_1}\beta_{t_2}|R\rangle|\uparrow\downarrow\rangle_{12}\nonumber\\
&+&\alpha_c\beta_{t_1}\alpha_{t_2}|R\rangle|\downarrow\uparrow\rangle_{12}
+\alpha_c\beta_{t_1}\beta_{t_2}|R\rangle|\downarrow\downarrow\rangle_{12},\nonumber\\
\vert\Omega_0''\rangle&=&\beta_c\alpha_{t_1}\alpha_{t_2}|L\rangle|\uparrow\uparrow\rangle_{12}
+\beta_c\alpha_{t_1}\beta_{t_2}|L\rangle|\uparrow\downarrow\rangle_{12}\nonumber\\
&+&\beta_c\beta_{t_1}\alpha_{t_2}|L\rangle|\downarrow\uparrow\rangle_{12}
+\beta_c\beta_{t_1}\beta_{t_2}|L\rangle|\downarrow\downarrow\rangle_{12}.
\end{eqnarray}

Our scheme for  a three-qubit Fredkin gate (see Fig.\ref{Fig4})
consists of three parts (three rounds).

(i) The injecting photon is split into two wave-packets by PBS$_1$,
that is, the part in the state $|R\rangle$ and that in the state
$|L\rangle$. The photon in the state $|R\rangle$ does not pass
through the cavities, while the photon in the state $|L\rangle$ is
injected into the cavities. Before the photon in the state $\vert
L\rangle$ reaches PBS$_2$, the state of the three-qubit hybrid
system given by Eq.(\ref{eq.21}) changes to be
\begin{eqnarray}                          \label{eq.23}
\vert \Omega_1\rangle &=&\vert \Omega_0'\rangle+\vert
\Omega_1''\rangle,
\end{eqnarray}
with
\begin{eqnarray}                          \label{eq.24}
\vert\Omega_1''\rangle&=&\beta_c\alpha_{t_1}\alpha_{t_2}|L^\uparrow\rangle|\uparrow\uparrow\rangle_{12}
+\beta_c\alpha_{t_1}\beta_{t_2}|L^\uparrow\rangle|\uparrow\downarrow\rangle_{12}\nonumber\\
&+&\beta_c\beta_{t_1}\alpha_{t_2}|L^\uparrow\rangle|\downarrow\uparrow\rangle_{12}
+\beta_c\beta_{t_1}\beta_{t_2}|L^\uparrow\rangle|\downarrow\downarrow\rangle_{12}.\nonumber\\
\end{eqnarray}
Since the photon in the state $|R\rangle$ does not pass through the
cavities, $|\Omega_0'\rangle$ stays the same all the time and only
$|\Omega_0''\rangle$ is changed. The nonlinear interaction between
the photon and the QD inside cavity 1 makes $\vert\Omega_1''\rangle$
be changed as
\begin{eqnarray}            \label{eq.25}
|\Omega_2''\rangle
&=&-\beta_c\alpha_{t_1}\alpha_{t_2}|L^\uparrow\rangle|\uparrow\uparrow\rangle_{12}
-\beta_c\alpha_{t_1}\beta_{t_2}|L^\uparrow\rangle|\uparrow\downarrow\rangle_{12}\nonumber\\
&+&\beta_c\beta_{t_1}\alpha_{t_2}|R^\downarrow\rangle|\downarrow\uparrow\rangle_{12}
+\beta_c\beta_{t_1}\beta_{t_2}|R^\downarrow\rangle|\downarrow\downarrow\rangle_{12}.\nonumber\\
\end{eqnarray}
After the photon interacts with the QD inside cavity 1, the emitting
photon emerges in  spatial mode 5.  Next, the photon coming from
path 5 is injected into   cavity 2. After the photon interacts with
the QD inside cavity 2, $\vert\Omega_2''\rangle$ becomes
\begin{eqnarray}            \label{eq.26}
|\Omega_3''\rangle &=&
\beta_c\alpha_{t_1}\alpha_{t_2}|L^\uparrow\rangle|\uparrow\uparrow\rangle_{12}
-\beta_c\alpha_{t_1}\beta_{t_2}|R^\downarrow\rangle|\uparrow\downarrow\rangle_{12}\nonumber\\
&-&\beta_c\beta_{t_1}\alpha_{t_2}|R^\downarrow\rangle|\downarrow\uparrow\rangle_{12}
+\beta_c\beta_{t_1}\beta_{t_2}|L^\uparrow\rangle|\downarrow\downarrow\rangle_{12}.\nonumber\\
\end{eqnarray}
After the photon interacts with the QD inside cavity 2, it is
emitted from  path $8$ whether it is in the state
$|R^\downarrow\rangle$ and comes from  path 7 or it is in the state
$|L^\uparrow\rangle$ and comes from  path 6. Therefore, after the
first round, the state of the whole system becomes
\begin{eqnarray}            \label{eq.27}
|\Omega_2\rangle &=&|\Omega_1'\rangle+|\Omega_3''\rangle,\nonumber\\
&=&\alpha_c\alpha_{t_1}\alpha_{t_2}|R\rangle|\uparrow\uparrow\rangle_{12}+
\alpha_c\alpha_{t_1}\beta_{t_2}|R\rangle|\uparrow\downarrow\rangle_{12}\nonumber\\
&+&\alpha_c\beta_{t_1}\alpha_{t_2}|R\rangle|\downarrow\uparrow\rangle_{12}
+\alpha_c\beta_{t_1}\beta_{t_2}|R\rangle|\downarrow\downarrow\rangle_{12}\nonumber\\
&+&\beta_c\alpha_{t_1}\alpha_{t_2}|L^\uparrow\rangle|\uparrow\uparrow\rangle_{12}
-\beta_c\alpha_{t_1}\beta_{t_2}|R^\downarrow\rangle|\uparrow\downarrow\rangle_{12}\nonumber\\
&-&\beta_c\beta_{t_1}\alpha_{t_2}|R^\downarrow\rangle|\downarrow\uparrow\rangle_{12}
+\beta_c\beta_{t_1}\beta_{t_2}|L^\uparrow\rangle|\downarrow\downarrow\rangle_{12}.\nonumber\\
\end{eqnarray}

Combing Eqs.(\ref{eq.21})and (\ref{eq.26})  and Fig.\ref{Fig4}, one
can find that the effect of  this round can be described by a
unitary matrix $U$,
\begin{eqnarray}                                      \label{28}
U=\left(\begin{array}{cccccccc}
1&0&0&0&0&0&0&0\\
0&0&0&0&0&-1&0&0\\
0&0&0&0&0&0&-1&0\\
0&0&0&1&0&0&0&0\\
0&0&0&0&1&0&0&0\\
0&-1&0&0&0&0&0&0\\
0&0&-1&0&0&0&0&0\\
0&0&0&0&0&0&0&1\\
\end{array}\right),
\end{eqnarray}
in the  basis
$\{$$|R^\downarrow\rangle|\uparrow\uparrow\rangle_{12}$,
$|R^\downarrow\rangle|\uparrow\downarrow\rangle_{12}$,
$|R^\downarrow\rangle|\downarrow\uparrow\rangle_{12}$,
$|R^\downarrow\rangle|\downarrow\downarrow\rangle_{12}$,
$|L^\uparrow\rangle|\uparrow\uparrow\rangle_{12}$,
$|L^\uparrow\rangle|\uparrow\downarrow \rangle_{12}$,
$|L^\uparrow\rangle|\downarrow\uparrow \rangle_{12}$,
$|L^\uparrow\rangle|\downarrow\downarrow \rangle_{12}$$\}$  as
$|\Omega_3''\rangle=U|\Omega_0''\rangle$ and $\Omega_0'$ keeps the
same one all the time.

(ii) We lead  the photon emitting from  spatial mode 8  back to
cavity 1 by using the optical switches $S_1$ and $S_2$ (dashed
line).  After the second round, the photon emitting from   spatial
mode 8 is led back to cavity 1 again for the third round.  Before
and after the second round,  an $H_p$ operation (i.e., an HWP in
path 9) is performed on the photon emitting from  spatial mode 8,
and  an $H_e$ operation is also performed on each of  the electrons
in cavities 1 and 2 . Hence, before the third round, the state of
the whole system is changed to be
\begin{eqnarray}            \label{eq.29}
|\Omega_3\rangle &=&|\Omega_1'\rangle\nonumber\\
&+&((H_p \otimes H_{e}\otimes H_{e})\cdot U \cdot (H_p \otimes H_{e}\otimes H_{e}))|\Omega_3''\rangle,\nonumber\\
&=&\alpha_c\alpha_{t_1}\alpha_{t_2}|R\rangle|\uparrow\uparrow\rangle_{12}+
\alpha_c\alpha_{t_1}\beta_{t_2}|R\rangle|\uparrow\downarrow\rangle_{12}\nonumber\\
&+&\alpha_c\beta_{t_1}\alpha_{t_2}|R\rangle|\downarrow\uparrow\rangle_{12}
+\alpha_c\beta_{t_1}\beta_{t_2}|R\rangle|\downarrow\downarrow\rangle_{12}\nonumber\\
&+&\beta_c\alpha_{t_1}\alpha_{t_2}|L^\uparrow\rangle|\uparrow\uparrow\rangle_{12}
-\beta_c\alpha_{t_1}\beta_{t_2}|R^\downarrow\rangle|\downarrow\uparrow\rangle_{12}\nonumber\\
&-&\beta_c\beta_{t_1}\alpha_{t_2}|R^\downarrow\rangle|\uparrow\downarrow\rangle_{12}
+\beta_c\beta_{t_1}\beta_{t_2}|L^\uparrow\rangle|\downarrow\downarrow\rangle_{12}.\nonumber\\
\end{eqnarray}

(iii) After the third round, the state of the whole system becomes
\begin{eqnarray}            \label{eq.30}
|\Omega_4\rangle &=&|\Omega_1'\rangle\nonumber\\
&+&
U(\beta_c\alpha_{t_1}\alpha_{t_2}|L^\uparrow\rangle|\uparrow\uparrow\rangle_{12}
-\beta_c\alpha_{t_1}\beta_{t_2}|R^\downarrow\rangle|\downarrow\uparrow\rangle_{12}\nonumber\\
&-&\beta_c\beta_{t_1}\alpha_{t_2}|R^\downarrow\rangle|\uparrow\downarrow\rangle_{12}
+\beta_c\beta_{t_1}\beta_{t_2}|L^\uparrow\rangle|\downarrow\downarrow\rangle_{12}),\nonumber\\
&=&
\alpha_c\alpha_{t_1}\alpha_{t_2}|R\rangle|\uparrow\uparrow\rangle_{12}+
\alpha_c\alpha_{t_1}\beta_{t_2}|R\rangle|\uparrow\downarrow\rangle_{12}\nonumber\\
&+&\alpha_c\beta_{t_1}\alpha_{t_2}|R\rangle|\downarrow\uparrow\rangle_{12}
+\alpha_c\beta_{t_1}\beta_{t_2}|R\rangle|\downarrow\downarrow\rangle_{12}\nonumber\\
&+&\beta_c\alpha_{t_1}\alpha_{t_2}|L^\uparrow\rangle|\uparrow\uparrow\rangle_{12}
+\beta_c\alpha_{t_1}\beta_{t_2}|L^\uparrow\rangle|\downarrow\uparrow\rangle_{12}\nonumber\\
&+&\beta_c\beta_{t_1}\alpha_{t_2}|L^\uparrow\rangle|\uparrow\downarrow\rangle_{12}
+\beta_c\beta_{t_1}\beta_{t_2}|L^\uparrow\rangle|\downarrow\downarrow\rangle_{12}.\nonumber\\
\end{eqnarray}

When the two wavepackets from  spatial modes 1 and 8 pass through
PBS$_4$ simultaneously, the system composed of the photon and the
two electrons is in the state
\begin{eqnarray}            \label{eq.31}
|\Omega_{5}\rangle &=&
\alpha_c|R\rangle(\alpha_{t_1}|\uparrow\rangle +
\beta_{t_1}|\downarrow\rangle)_{1}(\alpha_{t_2}|\uparrow\rangle +
\beta_{t_2}|\downarrow\rangle)_{2}
\nonumber\\
&+&\beta_c|L\rangle(\alpha_{t_2}|\uparrow\rangle +
\beta_{t_2}|\downarrow\rangle)_{1}(\alpha_{t_1}|\uparrow\rangle +
\beta_{t_1}|\downarrow\rangle)_{2}.\nonumber\\
\end{eqnarray}

From Eq.(\ref{eq.31}), one can see that the states of the two
solid-state target qubits (the two electron spins in cavities 1 and
2) are swapped when the photon qubit is in the state $\vert
L\rangle$, while they do not swap when the photon qubit is in the
state $\vert R\rangle$. The quantum circuit shown in Fig.\ref{Fig4}
can be used to construct the Fredkin gate on a three-qubit hybrid
system in a deterministic way.

\section{ Fidelities and efficiencies of the gates}\label{Sec6}

A crucial component in  our schemes for deterministic hybrid quantum
gates is the $X^--$cavity system. The quantum circuits
aforementioned for hybrid quantum gates are all under the ideal case
in which the success probability for each gate is 100\% in
principle.  Here, we assume that the Hadamard operation on the
electron is perfect as the spin manipulation technique is well
developed by using pulsed magnetic-resonance, nanosecond microwave
pulse, or picosecond/femtosecond optical pulse techniques
\cite{manipulating1,manipulating2,manipulating3,manipulating4}. The
optical elements, such as PBS, HWP, and optical switches, are also
assumed to be perfect, that is, the yield of the photon is 100\%. In
a realistic $X^--$cavity system, the side leakage should be taken
into account, and it affects the success probability of the device.
The whole process can be represented by the Heisenberg equations of
motion and the input-output relations \cite{Heisenberg}
\begin{eqnarray}   \label{eq.32}
\frac{d \hat{a}}{d t}&=&-\left[ i(\omega_c-\omega)+\kappa+\frac{\kappa_s}{2} \right]\hat{a}-g\sigma_-\nonumber\\
&&-\sqrt{\kappa}\,\hat{a}_{in}-\sqrt{\kappa}\,\hat{a}_{in}'+\hat{H},\nonumber\\
\frac{d \sigma_-}{d t}&=&-\left[ i(\omega_{X^-}-\omega)+\frac{\gamma}{2} \right]\sigma_--g\sigma_z\hat{a}+\hat{G},\nonumber \\
\hat{a_r}&=&\hat{a}_{in}+\sqrt{\kappa}\,\hat{a},\nonumber\\
\hat{a_t}&=&\hat{a}_{in}'+\sqrt{\kappa}\,\hat{a}.
\end{eqnarray}
Here $\hat{a}$ and $\sigma_-$  are the cavity field operator and the
$X^-$ dipole operator, respectively. $\hat{H}$ and $\hat{G}$ are the
noise operators related to the reservoirs. $\hat{a}_{in}$ and
$\hat{a}_{in}'$ are the two input field operators, shown in
Fig.\ref{Fig1}. $\hat{a}_r$ and $\hat{a}_t$ are the two output field
operators. $\omega$, $\omega_c$, and $\omega_{X^-}$ are the
frequencies of the input photon, the cavity mode, and the $X^-$
transition, respectively. $\gamma/2$ and  $\kappa$ ($\kappa_s/2$)
are the dipole decay rate and the cavity field decay rate (the side
leakage rate), respectively. $g$ is the coupling strength between
$X^-$ and the cavity mode.  By taking a weak approximation, the
reflection and the transmission coefficients of the  $X^--$cavity
system $r(\omega)$ and $t(\omega)$  can be obtained as \cite{Hu2}
\begin{eqnarray}    \label{eq.33}
r(\omega)&=&1+t(\omega),\nonumber\\
t(\omega)&=&\frac{-\kappa[i(\omega_{X^{-}}-\omega)+\frac{\gamma}{2}]}{[i(\omega_{X^{-}}
-\omega)
+\frac{\gamma}{2}][i(\omega_c-\omega)+\kappa+\frac{\kappa_s}{2}]+g^2}.\nonumber\\
\end{eqnarray}

In our works, we consider $\omega_{X^-}=\omega_c=\omega$, and the
reflection and the transmission coefficients of the uncoupled cavity
(cold cavity, described with the subscript $0$) and the coupled
cavity (hot cavity) can be simplified as
\begin{eqnarray}  \label{eq.34}
r_0(\omega)=\frac{\frac{\kappa_s}{2}}{\kappa+\frac{\kappa_s}{2}},\;\;\;\;
t_0(\omega)=-\frac{\kappa}{\kappa+\frac{\kappa_s}{2}},
\end{eqnarray}
and
\begin{eqnarray}  \label{eq.35}
r(\omega)=1+t(\omega),\;\;\;\;
t(\omega)=-\frac{\frac{\gamma}{2}\kappa}{\frac{\gamma}{2}[\kappa+\frac{\kappa_s}{2}]+g^2},
\end{eqnarray}
respectively. The rules for optical transitions in a realistic
$X^--$cavity system system become \cite{Hu2,CNOT}
\begin{eqnarray}    \label{eq.36}
|R^\downarrow\downarrow\rangle&\rightarrow&|r||L^\uparrow\downarrow\rangle+|t||R^\downarrow\downarrow\rangle,\nonumber\\
|L^\uparrow\downarrow\rangle&\rightarrow&|r||R^\downarrow\downarrow\rangle+|t||L^\uparrow\downarrow\rangle,\nonumber\\
|R^\downarrow\uparrow\rangle&\rightarrow&-|t_0||R^\downarrow\uparrow\rangle-|r_0||L^\uparrow\uparrow\rangle,\nonumber\\
|L^\uparrow\uparrow\rangle&\rightarrow&-|t_0||L^\uparrow\uparrow\rangle-|r_0||R^\downarrow\uparrow\rangle.
\end{eqnarray}
From Eq.(\ref{eq.36}), one can find that  the reflection and the
transmission coefficients connect to the fidelities and efficiencies
of our universal quantum gates. Substituting Eq.(\ref{eq.1}) with
Eq.(\ref{eq.36}), and combing the arguments made in Sec.\ref{Sec3},
we find that  the state of the system described by Eq.(\ref{eq.8})
becomes
\begin{eqnarray}    \label{eq.37}
\vert \Psi_4'\rangle
&=&\alpha_c\alpha_t|R\rangle|\uparrow\rangle+\alpha_c\beta_t|R\rangle|\downarrow\rangle\nonumber\\
&+&\underline{\frac{\beta_c\alpha_t}{2}(|t_0|+|r_0|-|t|-|r|)|L\rangle|\uparrow\rangle}\nonumber\\
&+&\frac{\beta_c\alpha_t}{2}(|t_0|+|r_0|+|t|+|r|)|L\rangle|\downarrow\rangle\nonumber\\
&+&\frac{\beta_c\beta_t}{2}(|t_0|+|r_0|+|t|+|r|)|L\rangle|\uparrow\rangle\nonumber\\
&+&\underline{\frac{\beta_c\beta_t}{2}(|t_0|+|r_0|-|t|-|r|)|L\rangle|\downarrow\rangle}.
\end{eqnarray}
The terms with underlines indicate the states which take the
bit-flip error.

Defining the fidelity of a quantum gate as
$F=|\langle\Psi_r|\Psi_i\rangle|^2$. Here $\vert \Psi_r\rangle$ and
$\vert \Psi_i\rangle$ are  the final states of the hybrid system in
our schemes for quantum gates in the realistic condition and the
ideal condition, respectively.  Therefore, the fidelity of our CNOT
gate on the two-qubit hybrid system discussed in Sec.\ref{Sec3} can
be written as
\begin{eqnarray}    \label{eq.38}
F_{CNOT}=\left[\frac{1+|t_0|+|r_0|}{2}\right]^2=1.
\end{eqnarray}
Similar calculations can be done for the Toffoli and the Fredkin
gates on the three-qubit hybrid system discussed in Secs. \ref{Sec4}
and  \ref{Sec5}, respectively.  That is,  the fidelities of the
Toffoli and the Fredkin gates $F_{T}$ and $F_{F}$ can be expressed
as
\begin{eqnarray}    \label{eq.39}
F_{T}&=&\frac{1}{16}
\large[2+|r_0|\left(|r|^2-|t|^2+|r_0|^2+|t_0|^2\right)\nonumber\\&&+|t_0|\left(1+|r_0|^2+|t_0|^2\right)
\large]^2,\\
F_{F} &=& \frac{1}{64} \large[4 + (1 +(|r_0|-|t|)
(|r|-|t_0|))\nonumber\\&\times& (|r|+|r_0|-|t|-|t_0|)
(|r|-|r_0|-|t|+|t_0|) \nonumber\\&+&
   2 (|r|+|r_0|) (|t|+|t_0|) \nonumber\\&+&
   \frac{1}{2} ((|r|-|t|)^2 + (|r_0|-|t_0|)^2) \nonumber\\&\times&((|r|+|r_0|)^2 +
   (|t|+|t_0|)^2)
\large]^2.
\end{eqnarray}

Defining the efficiency of a quantum gate as the ratio of the number
of the outputting photons to the inputting photons. The efficiency
is also determined by the reflection and the transmission
coefficients of the $X^--$cavity system. The efficiencies of these
gates can be written as
\begin{eqnarray}    \label{eq.41}
\eta_{CNOT}&=&\frac{1}{2}\left(1+\frac{|r|^2+|t|^2+|r_0|^2+|t_0|^2}{2}\right),\nonumber\\
\eta_{T}&=&\frac{1}{2}\left[1+\left(\frac{|r|^2+|t|^2+|r_0|^2+|t_0|^2}{2}\right)^3\right],\nonumber\\
\eta_{F}&=&\frac{1}{2}\left[1+\left(\frac{|r|^2+|t|^2+|r_0|^2+|t_0|^2}{2}\right)^6\right].\;\;\;\;\;
\end{eqnarray}

\begin{figure}[!h]
\begin{center}
\includegraphics[width=6.5 cm,angle=0]{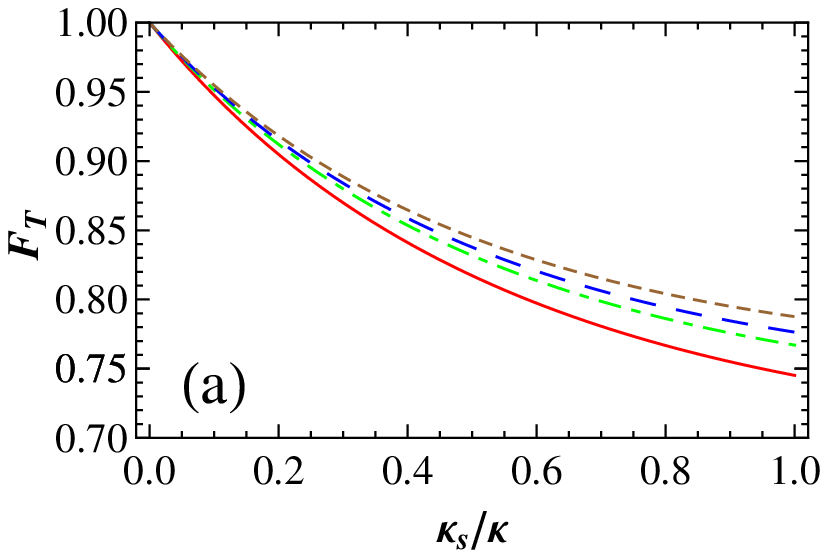}\\
\bigskip 
\includegraphics[width=6.5 cm,angle=0]{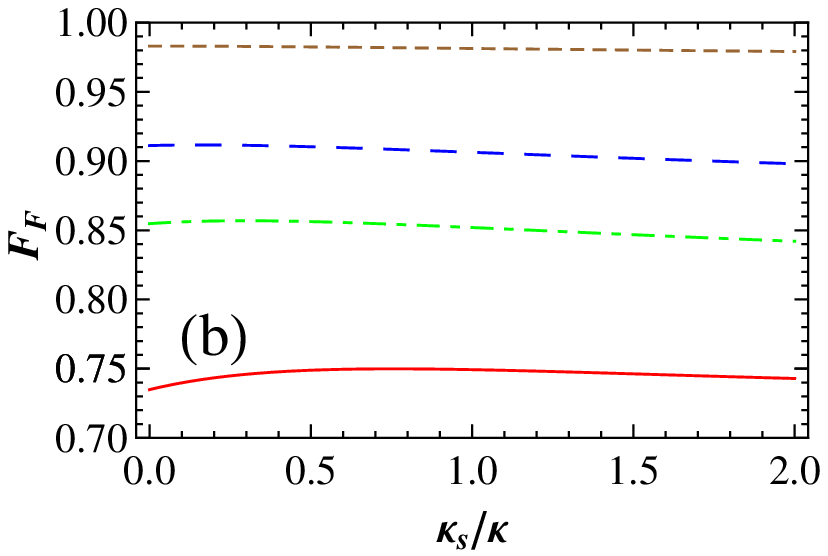}
\caption{(Color online) The fidelities of our three-qubit gates vs
the side leakage rate $\kappa_s/\kappa$ and the coupling strength
$g/\kappa$. (a) The fidelity of our Toffoli gate ($F_T$). (b) The
fidelity of our Fredkin gate ($F_F$). The solid (red), dashed-dotted
(green), large-dashed (blue), and dashed (brown) lines correspond to
$g=0.5\kappa$, $g=0.75\kappa$, $g=1.0\kappa$, and $g=2.4\kappa$,
respectively.
 $\gamma=0.1\kappa$ which is experimentally achievable and
$\omega_{X^-}=\omega_c=\omega$ are taken here.} \label{Fig5}
\end{center}
\end{figure}

It is still a big challenge to achieve  strong coupling in
experiments. However,   strong coupling has been demonstrated in
various microcavity- and nanocavity-QD systems recently
\cite{observed1,observed2,observed3,observed4,observed5}.
$g/(\kappa+\kappa_s)\simeq0.5$ for  micropillar microcavities with a
diameter of $1.5\,\mu m$  and a quality factor of $Q=8800$ have been
reported \cite{observed1}.  $Q$ is dominated by $\kappa$ and
$\kappa_s$. For micropillar systems, $Q$ can be increased to $\sim 4
\times 10^4$ ($g/(\kappa+\kappa_s)\simeq2.4$) by improving the
sample designs, growth, and fabrication \cite{observed4}. By taking
high-$Q$ micropillars and thinning down the top mirrors, $Q$ can
decrease to $\sim1.7\times10^4$ with $\kappa_s/\kappa=0.7$ and
$g/(\kappa+\kappa_s)\simeq 1.0$, and the system remains in the
strong coupling regime \cite{Hu4}. Recent experiments have indicated
that the strong coupling can be achieved for $d=7.3\,\mu m$
micropillars, where the side leakage is small compared with that of
small micropillars \cite{micropillars}.

\begin{figure}[tpb]    
\begin{center}
\includegraphics[width=6.5 cm,angle=0]{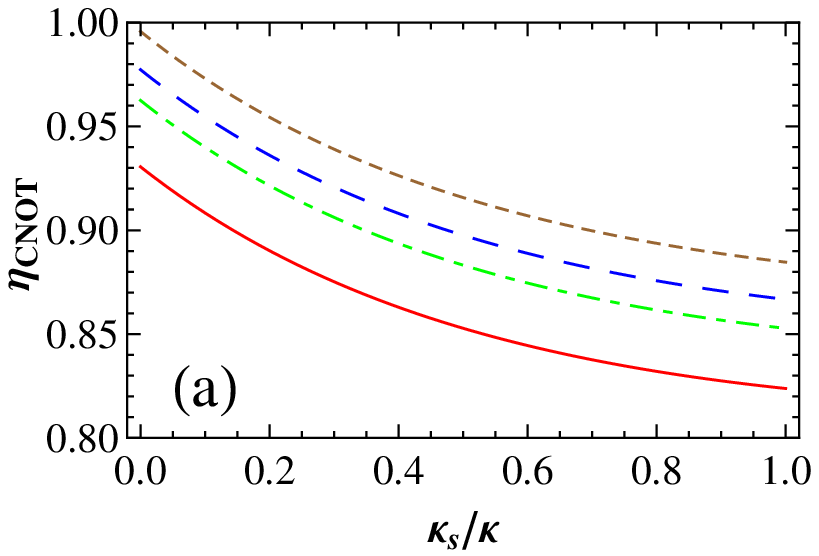}\\
\bigskip
\includegraphics[width=6.5 cm,angle=0]{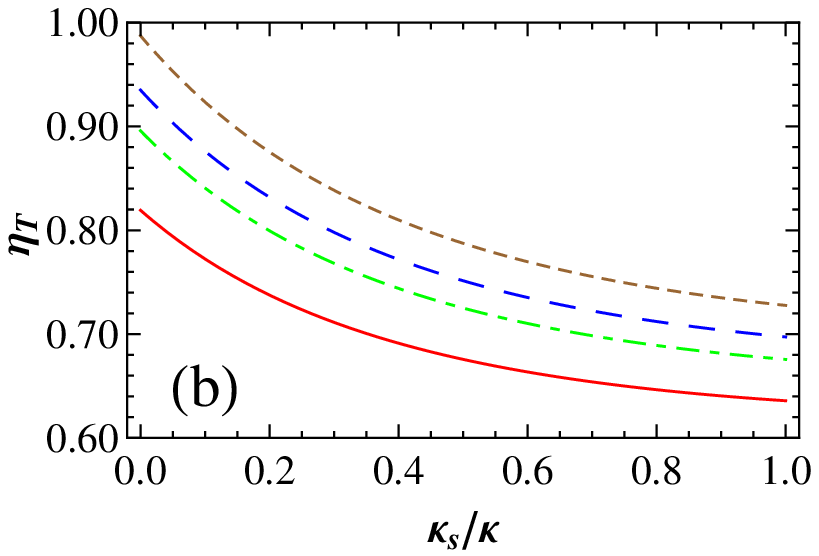}\\
\bigskip
\includegraphics[width=6.5 cm,angle=0]{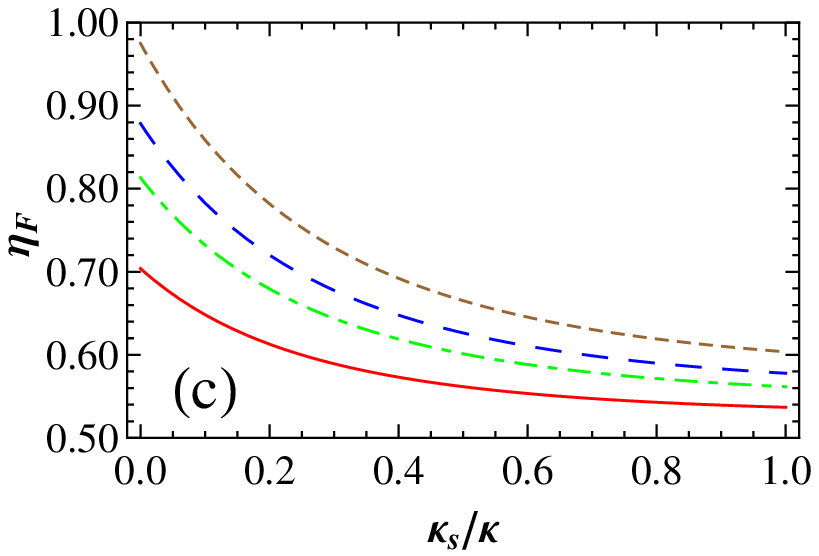}
\caption{(Color online)  The efficiencies of our universal quantum
gates vs the side leakage rate $\kappa_s/\kappa$ and the coupling
strength $g/\kappa$. (a) The efficiency of our CNOT gate
($\eta_{CNOT}$). (b) The efficiency of our Toffoli gate
($\eta_{T}$).  (c) The efficiency of our Fredkin gate ($\eta_{F}$).
The solid (red), dashed-dotted (green), large-dashed (blue), and
dashed (brown) lines correspond to $g=0.5\kappa$, $g=0.75\kappa$,
$g=1.0\kappa$, and $g=2.4\kappa$, respectively. We take
$\gamma=0.1\kappa$ and $\omega_{X^-}=\omega_c=\omega$ for panels
(a), (b), and (c).}\label{Fig6}
\end{center}
\end{figure}

The relations between the fidelities of our Toffoli gate or our
Fredkin gate and  the side leakage rate $\kappa_s/\kappa$ and the
coupling strength $g/\kappa$ are shown in Fig.\ref{Fig5}. The
efficiencies  of our universal quantum gates are shown in
Fig.\ref{Fig6}. For our schemes for hybrid quantum gates, in the
weak regime, if $g/\kappa=0.5$ and $\kappa_s/\kappa=0.25$,
$F_T=88.7\%$ and $F_F=74.5\%$ with $\eta_{CNOT}=88.2\%$,
$\eta_T=72.4\%$, $\eta_F=60\%$; and $F_T=100\%$ and $F_F=73.5\%$
with $\eta_{CNOT}=93.1\%$, $\eta_T=81.9\%$, and $\eta_F=70.4\%$ when
$\kappa_s/\kappa=0$. In the strong regime, when $g/\kappa=2.4$ and
$\kappa_s/\kappa=0.5$, $F_T=84.5\%$ and $F_F=98.2\%$ with
$\eta_{CNOT}=91.6\%$, $\eta_T=78.8\%$, and $\eta_F=66.5\%$. If the
cavity leakage can be neglected, both the fidelity and the
efficiency can reach near-unity ($F_{T}=100\%$ and $F_F=98.3\%$ with
$\eta_{CNOT}=99.6\%$, $\eta_T=98.7\%$, and $\eta_F=97.5\%$).
$F_T=80.6\%$ and $F_F=90.9\%$ with $\eta_{CNOT}=88.2\%$,
$\eta_T=72.2\%$, and $\eta_F=59.9\%$ when $g/\kappa=1.0$ and
$\kappa_s/\kappa=0.7$ ($F_T=100\%$ and $F_F=91.1\%$ with
$\eta_{CNOT}=97.7\%$, $\eta_T=93.5\%$, and $\eta_F=87.8\%$ when
$\kappa_s/\kappa=0$). It is worth   pointing out that $g/\kappa$ and
$\kappa_s/\kappa$ does not affect the fidelity of our CNOT gate and
it remains at unity.

Note that there are two kinds of exciton dephasing in the QD, the
optical dephasing, and the spin dephasing of $X^-$,  caused by the
exciton decoherence. Exciton dephasing reduces the fidelity by the
amount of
\begin{eqnarray}    \label{eq.42}
[1-\exp(-\tau/T_2)],
\end{eqnarray}
where $\tau$ and $T_2$  are the cavity photon lifetime and the trion
coherence time, respectively. The optical dephasing reduces the
fidelity less than 10\% because the time scale of the excitons can
reach hundreds of picoseconds
\cite{opticl-deph1,opticl-deph2,opticl-deph3}, while the cavity
photon lifetime is in the tens of picoseconds range for a
self-assembled In(Ga)As-based QD with a cavity \emph{Q} factor  of
$10^4 -10^5$ in the strong coupling regime. The effect of the spin
dephasing can be neglected because the spin decoherence time
($T_2^h>100 ns$) is several orders of  magnitude longer than the
cavity photon lifetime (typically $\tau<10 ps$)
\cite{spin-deph1,spin-deph2,spin-deph3}.

In a realistic QD [e.g., for self-assembled In(Ga)As QDs], imperfect
optical selection induced by heavy-light hole mixing reduces the
fidelity by a few percent \cite{optic-selec}. However, this
undesirable mixing can be reduced by engineering the shape and the
size of QDs or choosing the types of QDs.

\section{Discussion and summary}\label{Sec7}

We have proposed some schemes for implementing deterministic
universal quantum logic gates between flying photon qubits and
stationary electron-spin qubits with linear optical elements.
Different from the work in Ref.\cite{CNOT} which presents a scheme
for a CNOT gate with the confined electron as the control qubit and
the photon as the target qubit, the CNOT gate in our work takes the
flying photon as the control qubit and the electron as the target
qubit. Compared with the two-qubit case, the works for implementing
multi-qubit gates are  generally quite complex and difficult. In
this work, the schemes for constructing three-qubit universal
quantum gates (Toffoli and Fredkin) in hybrid systems are also
discussed.

The control qubit of the gates in our work is encoded on the photon
which is easy to be manipulated. The target qubits are encoded on
the electrons confined in QDs inside optical microcavities which can
be used for processor and quantum computation. It is worth
mentioning that our schemes require no additional qubits. This good
feature reduces not only quantum resources but also errors. Since
the electron-spin qubit is confined in a QD inside a double-sided
microcavity, our gates are robust. The gate based on the
single-sided QD-cavity system  demands the transmission coefficient
of the uncoupled cavity is balanceable with the reflection
coefficient of the coupled cavity to get a high fidelity \cite{Hu2}.

The cavity leakage and the cavity loss induce the bit-flip error and
the different transmittance or reflectance between the hot cavity
and the cold cavity in an $X^--$cavity system. These factors  reduce
the fidelity and the efficiency of our gates. We have shown that a
high fidelity and efficiency can be achieved in  both the weak
coupling regime and in  the strong coupling regime in our schemes.
If the cavity leakage is much lower than the cavity loss (the ideal
case), the fidelity and  the efficiency of our gates can reach
near-unity in the strong coupling regime.

\section*{ACKNOWLEDGMENTS}

This work is supported by the National Natural Science Foundation of
China under Grant Nos. 10974020 and 11174039,  NCET-11-0031, and the
Fundamental Research Funds for the Central Universities.


\begin{thebibliography}{99}

 \bibitem{book}  M. A. Nielsen and I. L. Chuang, \emph{Quantum Computation and Quantum Information} (Cambridge University Press, Cambridge, UK, 2000).


\bibitem{uni}  A. Barenco, C. H. Bennett, R. Cleve, D. P. DiVincenzo, N. Margolus, P. Shor, T. Sleator, J. A. Smolin,
and H. Weinfurter, Phys. Rev. A \textbf{52}, 3457 (1995).


\bibitem{2-qubit1} G. Vidal and C. M. Dawson, Phys. Rev. A \textbf{69}, 010301 (2004).


\bibitem{2-qubit2} F. Vatan and C. Williams, Phys. Rev. A \textbf{69}, 032315 (2004).


\bibitem{2-qubit3} V. V. Shende, I. L. Markov, and S. S. Bullock, Phys. Rev. A \textbf{69}, 062321 (2004).


\bibitem{2-qubit4} V. V. Shende, S. S. Bullock. I. L. Markov. Phys. Rev. A \textbf{70}, 012310 (2004).


\bibitem{Toffoli} Y. Y. Shi, Quantum Inf. Comput. \textbf{3}, 84 (2003).


\bibitem{Fredkin} E. Fredkin and T. Toffoli, Int. J. Theor. Phys. \textbf{21}, 219 (1982).


\bibitem{Shor} P. W. Shor, SIAM J. Sci. Stat. Comput. \textbf{26}, 1484 (1997).


\bibitem{error} D. G. Cory, M. D. Price, W. Maas, E. Knill, R. Laflamme, W. H. Zurek, T. F. Havel, and S. S. Somaroo,
Phys. Rev. Lett. \textbf{81}, 2152 (1998).


\bibitem{fault} E. Dennis, Phys. Rev. A \textbf{63}, 052314 (2001).


\bibitem{swap} L. M. Liang and C. Z. Li, Phys Rev. A \textbf{72}, 024303 (2005).


\bibitem{QD1} D. Loss and D. P. DiVincenzo, Phys. Rev. A \textbf{57}, 120 (1998).


\bibitem{QD2} A. Imamoglu, D. D. Awschalom, G. Burkard, D. P. DiVincenzo, D. Loss, M. Sherwin, and A. Small, Phys. Rev. Lett. \textbf{83}, 4204 (1999).


\bibitem{QD3} C. Piermarocchi, P. C. Chen, L. J. Sham, and D. G. Steel, Phys. Rev. Lett. \textbf{89}, 167402 (2002).


\bibitem{QD4} T. Calarco, A. Datta, P. Fedichev, E. Pazy, and P. Zoller, Phys. Rev. A \textbf{68}, 012310 (2003).


\bibitem{QD5} S. M. Clark, K. M. C. Fu, T. D. Ladd, and Y. Yamamoto, Phys. Rev. Lett. \textbf{99}, 040501 (2007).


\bibitem{QD6} Z. R. Lin, G. P. Guo, T. Tu, F. Y. Zhu, and G. C. Guo, Phys. Rev. Lett. \textbf{101}, 230501 (2008).



\bibitem{coher-time1} J. R. Petta, A. C. Johnson, J. M. Taylor, E. A. Laird, A. Yacoby, M. D. Lukin, C. M. Marcus, M. P. Hanson,
and A. C. Gossard, Science \textbf{309}, 2180 (2005).


\bibitem{coher-time2} A. Greilich, D. R. Yakovlev, A. Shabaev, A. L. Efros, I. A. Yugova, R. Oulton, V. Stavarache, D. Reuter, A. Wieck,
and M. Bayer, Science \textbf{313}, 341 (2006).


\bibitem{relaxation1} J. M. Elzerman, R. Hanson, L. H. Willems van Beveren, B. Witkamp, L. M. K. Vandersypen, and L. P. Kouwenhoven,
Nature (London)  \textbf{430}, 431 (2004).


\bibitem{relaxation2} M. Kroutvar, Y. Ducommun, D. Heiss, M. Bichler, D. Schuh, G. Abstreiter, and J. J. Finley, Nature (London) \textbf{432}, 81 (2004).


\bibitem{cooling1} M. Atat\"{u}re, J. Dreiser, A. Badolato, A. H\"{o}gele, K. Karrai, and A. Imamoglu, Science \textbf{312}, 551  (2006).


\bibitem{cooling2} X. D. Xu, Y. W. Wu, B. Sun, Q. Huang, J. Cheng, D. G. Steel, A. S. Bracker, D. Gammon, C. Emary, and L. J. Sham,
                    Phys. Rev. Lett. \textbf{99}, 097401  (2007).


\bibitem{manipulating1}  J. Berezovsky, M. H. Mikkelsen, N. G. Stoltz, L. A. Coldren, and D. D. Awschalom, Science  \textbf{320}, 349 (2008).


\bibitem{manipulating2}  D. Press, T. D. Ladd, B. Y. Zhang, and Y. Yamamoto, Nature (London) \textbf{456}, 218 (2008).


\bibitem{Hu1}  C. Y. Hu, A. Young, J. L. O'Brien, W. J. Munro, and J. G. Rarity, Phys. Rev. B \textbf{78}, 085307 (2008).


\bibitem{Hu2}  C. Y. Hu, W. J. Munro, J. L. O'Brien, and J. G. Rarity, Phys. Rev. B \textbf{80}, 205326 (2009).


\bibitem{CNOT}  C. Bonato, F. Haupt, S. S. R. Oemrawsingh, J. Gudat, D. Ding, M. P. van Exter, and D. Bouwmeester,
Phys. Rev. Lett. \textbf{104}, 160503 (2010).


\bibitem{Hu3}  C. Y. Hu, W. J. Munro, and J. G. Rarity, Phys. Rev. B \textbf{78}, 125318 (2008).


\bibitem{Hu4}  C. Y. Hu and J. G. Rarity, Phys. Rev. B \textbf{83}, 115303 (2011).


\bibitem{Appli1} C. Wang, Y. Zhang, and G. S. Jin, Phys. Rev. A \textbf{84}, 032307 (2011).


\bibitem{Appli2} T. Yu, A. D. Zhu, S. Zhang, K. H. Yeon, and S. C. Yu, Phys. Scr. \textbf{84}, 025001 (2011).


\bibitem{Appli3} T. J. Wang, S. Y. Song,  and  G. L. Long, Phys. Rev. A, \textbf{85}, 062311 (2012).


\bibitem{Renbaocang} B. C. Ren, H. R. Wei, M. Hua, T. Li, and F. G. Deng, Opt. Express \textbf{20}, 24664 (2012).


\bibitem{exciton1}  R. J. Warburton, C. S. D\"{u}rr, K. Karrai, J. P. Kotthaus, G. Medeiros-Ribeiro, and P. M. Petroff,
Phys. Rev. Lett. \textbf{79}, 5282 (1997).


\bibitem{exciton2} C. Y. Hu, W. Ossau, D. R. Yakovlev, and G. Landwehr, T. Wojtowicz, G. Karczewski, and J. Kossut,
Phys. Rev. B \textbf{58}, R1766 (1998).


\bibitem{cost}  V. V. Shende and I. L. Markov, Quant. Inf. Comput. \textbf{9}, 461 (2009).


\bibitem{manipulating3} J. A. Gupta, R. Knobel, N. Samarth, and D. D. Awschalom, Science \textbf{292}, 2458  (2001).


\bibitem{manipulating4} P. C. Chen, C. Piermarocchi, L. J. Sham, D. Gammon, and D. G. Steel, Phys. Rev. B \textbf{69}, 075320  (2004).


\bibitem{Heisenberg}  D. F. Walls and G. J. Milburn, \emph{Quantum Optics}  (Springer-Verlag, Berlin, 1994).


\bibitem{observed1}  J. P. Reithmaier, G. Sek, A. L\"{o}ffler, C. Hofmann, S. Kuhn, S. Reitzenstein, L. V. Keldysh, V. D. Kulakovskii,
 T. L. Reinecke, and A. Forchel, Nature (London) \textbf{432}, 197 (2004).


\bibitem{observed2}  T. Yoshie, A. Scherer, J. Hendrickson, G. Khitrova, H. M. Gibbs, G. Rupper, C. Ell, O. B. Shchekin,
and D. G. Deppe, Nature (London) \textbf{432}, 200 (2004).


\bibitem{observed3}  E. Peter, P. Senellart, D. Martrou, A. Lema\^{I}tre, J. Hours, J. M. G\'{e}rard, and J. Bloch,
Phys. Rev. Lett. \textbf{95}, 067401 (2005).


\bibitem{observed4}  S. Reitzenstein, C. Hofmann, A. Gorbunov, M. Strau{\ss}, S. H. Kwon, C. Schneider, A. L\"{o}ffler, S. H\"{o}fling,
M. Kamp, and A. Forchel, Appl. Phys. Lett. \textbf{90}, 251109 (2007).


\bibitem{observed5}  A. B. Young, R. Oulton, C. Y. Hu, A. C. T. Thijssen, C. Schneider, S. Reitzenstein, M. Kamp, S. H\"{o}fling,
L. Worschech, A. Forchel, and J. G. Rarity,  Phys. Rev. A  \textbf{84}, 011803 (2011).


\bibitem{micropillars}  V. Loo, L. Lanco, A. Lema\^{i}tre, I. Sagnes, O. Krebs, P. Voisin, and P. Senellart,
Appl. Phys. Lett. \textbf{97}, 241110 (2010).


\bibitem{opticl-deph1}  P. Borri, W. Langbein, S. Schneider, U. Woggon, R. L. Sellin, D. Ouyang, and D. Bimberg,
Phys. Rev. Lett. \textbf{87}, 157401 (2001).


\bibitem{opticl-deph2} D. Birkedal, K. Leosson, and J. M. Hvam, Phys. Rev. Lett. \textbf{87}, 227401 (2001).


\bibitem{opticl-deph3}W. Langbein, P. Borri, U. Woggon, V. Stavarache, D. Reuter, and A. D. Wieck, Phys. Rev. B \textbf{70}, 033301 (2004).


\bibitem{spin-deph1} D. Heiss, S. Schaeck, H. Huebl, M. Bichler, G. Abstreiter, J. J. Finley, D. V. Bulaev, and D. Loss,
Phys. Rev. B \textbf{76}, 241306 (2007).


\bibitem{spin-deph2} B. D. Gerardot, D. Brunner, P. A. Dalgarno, P. \"{O}hberg, S. Seidl, M. Kroner, K. Karrai, N. G. Stoltz, P. M. Petroff,
                     and R. J. Warburton, Nature (London) \textbf{451}, 441 (2008).


\bibitem{spin-deph3} D. Brunner,  B. D. Gerardot, P. A. Dalgarno, G. W\"{u}st, K. Karrai, N. G. Stoltz, P. M. Petroff,
and R. J. Warburton, Science \textbf{325}, 70 (2009).


\bibitem{optic-selec}  G. Bester, S. Nair, and A. Zunger, Phys. Rev. B \textbf{67}, 161306 (2003).



\end{thebibliography}
\end{document}